\documentclass{desyproc}
\usepackage{url, hyperref}
\begin{document}
\title{CDMS-II to SuperCDMS: WIMP search at a zeptobarn}

\author{{\slshape Tobias Bruch$^1$ for the CDMS Collaboration}\\[1ex]
$^1$University of Z\"urich, Winterthurerstr 190, 8057 Z\"urich, Switzerland\\}

\contribID{bruch\_tobias}

\desyproc{DESY-PROC-2009-05}
\acronym{Patras 2009} 
\doi  

\maketitle

\begin{abstract}
The Cryogenic Dark Matter search experiment (CDMS) employs low-temperature Ge and Si detectors to detect WIMPs via their elastic scattering of target nuclei. The last analysis with an germanium exposure of 397.8\,kg-days resulted in zero observed candidate events, setting an upper limit on the spin-independent WIMP-nucleon cross-section of 6.6 $\times$ 10$^{-44}$\,cm$^2$ (4.6 $\times$ 10$^{-44}$\,cm$^2$, when previous CDMS Soudan data is included) for a WIMP mass of 60\,GeV. The improvements in the surface event rejection capability for the current analysis with an germanium exposure about a factor of 2.5 greater than used in the last analysis will be discussed. To increase the sensitivity beyond the 1$\times$10$^{-44}$\,cm$^2$ benchmark new 1 inch thick detectors have been developed. A first tower consisting of six of these detectors has been successfully installed at the Soudan site. These detectors will be used in a 15 kg SuperCDMS stage with an expected sensitivity on the spin-independent WIMP-nucleon elastic scattering cross-section of 5 $\times$10$^{-45}$\,cm$^2$. In addition, the CDMS Collaboration has started to look for signatures of non WIMP dark matter particles, which may explain the annual modulation signature observed by DAMA.
\end{abstract}
\section{Introduction}
The Cryogenic Dark Matter Search (CDMS) experiment operates 19 Ge ($\sim$\,250\,g each) and 11 Si ($\sim$\,100\,g each) detectors at the Soudan underground laboratory (MN, USA) to search for non-luminous, non-baryonic Weakly Interacting Massive Particles (WIMPs), that could form the majority of the matter in the universe \cite{Spergel,Jungman}. The detectors are designed to read out both ionization and phonon signals of an interaction. The ratio of ionization to phonon energy, the ionization yield, enables discrimination of nuclear recoils from electron recoils. The details of the detector structure and operation can be found in \cite{cdms2005} . The ionization yield discriminator provides a rejection factor of $>$\,10$^4$ for electron recoils, leaving surface events as the main background in the search for nuclear recoils.

\section{Surface contamination of the crystals}

Particle interactions may suffer from a suppressed ionization signal if the interactions occur in the first few microns of the crystal surfaces, this ionization loss is sufficient to missclassify such events as nuclear recoils. These surface events mainly occur due to radioactive contamination on detector surfaces, or as a result of external photon interactions releasing low-energy electrons from surfaces near the detectors. A correlation analysis between alpha-decay and surface-event rates provides evidence that $^{210}$Pb is a major component of the surface event background \cite{pb210background}. The correlation analysis also shows some indication that  improved detector handling during production and testing reduced the surface contamination for the later three towers. In table \ref{tab:210pbcounts} the measured surface event rates are summarized. The third row gives the remaining (non $^{210}$Pb related) surface-event rate which is compatible with the rate expected from photon induced events, given in the last row.

\begin{wraptable}{r}{0.55\textwidth}
\centerline{\begin{tabular}{|c|c|}
\hline
Surface event rate & 10-100 keV singles \\
  10$^{-3}$counts/detector/day & \\
\hline
Total observed & 371 $\pm$ 183 \\
$^{210}$Pb corr. analysis & 240 $\pm$ 183 \\
non $^{210}$Pb & 131 $\pm$ 63 \\
exp. photon induced & 217 $\pm$ 103 \\
\hline
            \end{tabular}
}
\caption{Surface event rate from $^{210}$Pb contamination and photon induced events.}
\label{tab:210pbcounts}
\end{wraptable}

To discriminate surface events against nuclear recoils the timing properties of the phonon pulses are used. Two possible parameters are the delay of the slower phonon signal with respect to the ionization signal and the risetime of the leading phonon pulse (which is the one with the highest amplitude), since surface events have smaller delays and faster risetimes than bulk nuclear-recoils. The left panel of figure \ref{fig:timingsensitivity} shows a comparison of the distribution of surface events and nuclear recoils in the simple timing discriminator (sum of delay and risetime). An improvement in the separation of the distributions of surface events and nuclear recoils may yield a high selection efficiency for nuclear recoils in the current analysis. In the current analysis new additional parameters have been defined which may provide an even higher selection efficiency of nuclear recoils while maintaining a high rejection of surface events.

\section{Recent results and the road to the zeptobarn sensitivity}
\begin{figure}[]

\begin{minipage}[c]{0.495\textwidth}
      \centering
\includegraphics[scale=0.38]{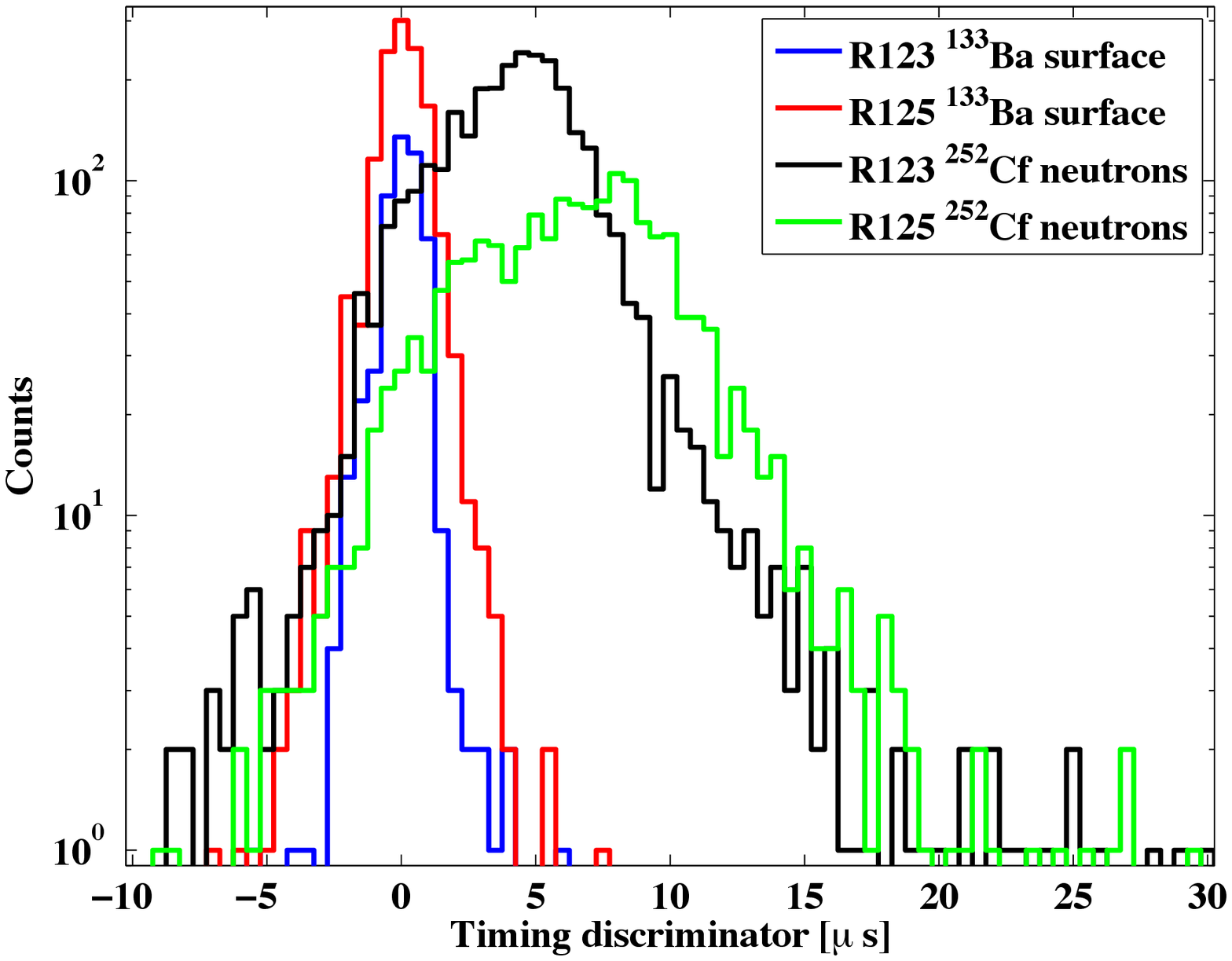}
 \end{minipage}
\begin{minipage}[c]{0.495\textwidth}
      \centering
\includegraphics[scale=0.73]{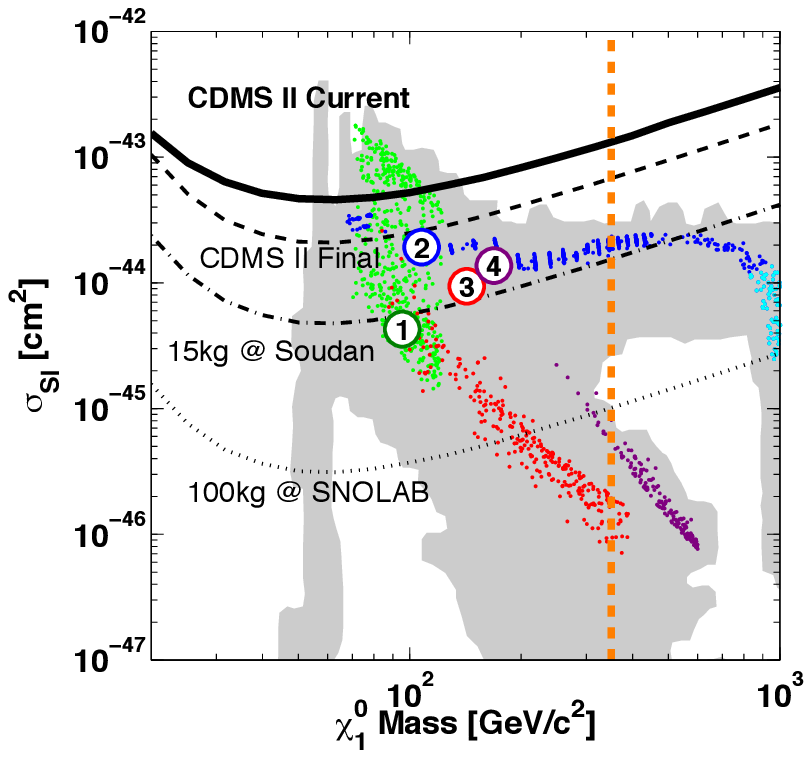}
 \end{minipage}
\caption{\small{Left panel: Comparison of the simple timing discriminator for the last analysis with the current analysis. The separation of the surface event distribution and nuclear recoil distribution shows an improvement for the current data, yielding a possible higher selection efficiency for nuclear recoils. Right panel: Mass vs spin-independent WIMP nucleon cross section parameter space. The horizontal curves represent current and projected sensitivities of the CDMS experiment. The vertical (orange/dotted) line represent the approximate upper limit of the LHC reach in neutralino mass. The gray shaded region and colored points are scans of the CMSSM \cite{baltz} along with four Linear Collider Cosmology benchmark points (numbered points)\cite{baltz2006}.}}\label{fig:timingsensitivity}
\end{figure}

The recent result used a germanium exposure of  397.8\,kg-days. Surface events present in $^{133}$Ba calibration data or naturally present in WIMP search data, were studied to determine the surface event leakage into the signal region after the timing cut is applied. The estimated surface event leakage, based on the observed numbers of single- and multiple- scatter events within and sourrounding the 2$\sigma$ nuclear recoil region in each detector, is $0.6^{+0.5}_{-0.3} (stat.) ^{+0.3}_{-0.2} (syst.)$ events \cite{r123analysis}. Upon the unblinding of the data no event was observed within the signal region. From this data the 90\% CL upper limit on the spin-independent WIMP-nucleon cross section is derived \cite{r123analysis}. The inclusion of a reanalysis of previous CDMS data, sets the world's most stringent upper limit on the spin-independent WIMP-nucleon cross section for WIMP masses above 42\,GeV/c$^2$ with a minimum of 4.6\,$\times$\,$10^{-44}$\,cm$^2$ for a WIMP mass of 60\,GeV/c$^2$ (shown as the black/solid line in figure \ref{fig:timingsensitivity}).

To further increase the sensitivity the total accumulated exposure (runtime times mass) has to be increased and the backround have to be keep under control. So far the CDMS-II setup has acquired an additional Ge exposure which is about a factor of 2.5 of the exposure used for the recent result. With this accumulated exposure the CDMS-II setup is expected to reach a sensitivity in the low 10$^{-44}$\,cm$^2$ range (see figure \ref{fig:timingsensitivity}). For the SuperCDMS setup new 1 inch thick detectors have been developed and tested, providing an increase of a factor 2.54 in mass with respect to the 1\,cm thick detectors used in CDMS-II. The first of 5 SuperTowers being operated at the Soudan site has been successfully installed and its performance in terms of backround rejection capability is currently investigated. The redesign of the phonon readout, which maximizes the active phonon collection area, and new sensor configurations are expected to improve the discrimination between surface events and nuclear-recoils. With a runtime of 2 years the five SuperTower setup with a total mass of 15kg of Ge is expected to break the benchmark sensitivity of 1\,$\times$\,$10^{-45}$cm$^{2}$ and probe parameter space down to a level of 5\,$\times$\,$10^{-45}$cm$^{2}$. At the SuperCDMS 100\,kg stage, operated at SNOLAB, it is expected that the CDMS experiment reaches the zeptobarn sensitivity. As shown in figure \ref{fig:timingsensitivity}, SuperCDMS 100\,kg aims to reach a sensitivity of 3\,$\times$\,$10^{-46}$\,cm$^{2}$ at a WIMP mass of 60\,GeV/c$^2$.

\section{Electromagnetic signatures of dark matter}
\begin{wrapfigure}{l}{0.5\textwidth}
\centering
\includegraphics[scale=0.4]{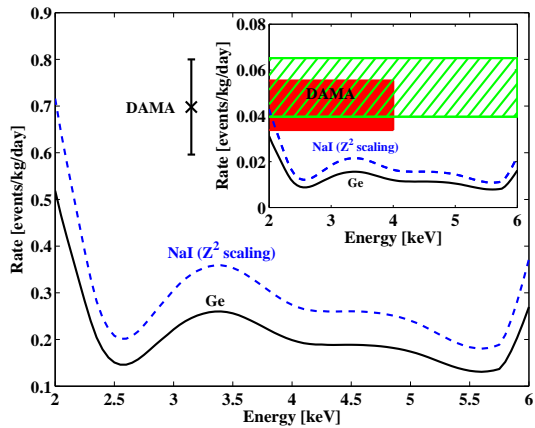}
\caption{\small{Direct comparison between the 90\% CL upper limit in CDMS (black/solid) with the total counting rate observed by DAMA/LIBRA (black data point, shown with 2$\sigma$ error bars). The inset compares the upper limit on the modulation, amplitude assumed to be 6\% of the upper limit on the total rate, with the 2$\sigma$ regions of the annual modulation amplitude observed by DAMA. The blue lines give the upper limits in Ge scaled to NaI (see text). ~\newline }}
\label{fig:damacomp}
\end{wrapfigure}

The annual modulation signature observed by DAMA \cite{damalibra} may be interpreted as the conversion of a dark matter particle into electromagnetic energy in the detector. In this case the corresponding signal should also be observable in the electron-recoil spectrum of CDMS. The possibility of an electron-recoil signal from axion-like dark matter particles has recently been investigated \cite{cdmsaxion}. A general analysis of the low-energy electron recoil specrum of the germanium detectors from 2-8.5 keV resulted in 90\% CL upper limits on an excess rate above background \cite{cdmslowen}. These upper limits are directly compared to the total rate observed by DAMA in figure \ref{fig:damacomp}. It should be stressed that the DAMA rate may contain a contribution of $^{40}$K decays at an energy of 3.2 keV, but no information on the actual rate from this background is provided by the DAMA collaboration. Thus no subtraction is performed, which would reduce the difference between the upper limit from CDMS and the excess rate in DAMA. The event rates in CDMS and DAMA detection media may differ depending on the coupling of the dark matter particle. For an electromagnetic conversion a Z$^2$ (where Z is the atomic number) scaling of the cross section is natural and thus considered here. Another scaling can be trivially considered. The scaled limits from CDMS to the rate in NaI are shown as the blue lines in figure \ref{fig:damacomp}. The inset in figure \ref{fig:damacomp} shows the upper limits on a possible modulation amplitude under the assumption of a standard halo model, yielding a conservative upper bound on the modulation amplitude of 6\% of the total rate. The upper limits set by CDMS are about a factor of 2 lower than the modulation amplitudes observed by DAMA. This constraints are not affected by any residual background rate in the DAMA data.

\section{Summary}
The CDMS-II experiment has maintained high dark matter discovery potential by limiting expected backgrounds to less than one event in the signal region. The current data sets the world's most stringent upper limit on the spin-independent WIMP-nucleon cross-section for WIMP masses above 42\,GeV/c$^2$ with a minimum of 4.6\,$\times $\,$10^{-44}$\,cm$^2$ for a WIMP mass of 60\,GeV/c$^2$. The analysis of a dataset with a factor of 2.5 more Ge exposure is ongoing, it is expected to increase the sensitivity to the low 10$^{-44}$\,cm$^2$ range. The first super tower with 6 new 1\,inch thick detectors was successfully installed at the Soudan site and is currently tested in terms of rejection efficiency of surface events. These detectors will be used in the next upgrades of the CDMS experiment which aims to reach the zeptobarn sensitivity with a SuperCDMS 100\,kg stage operated at SNOLAB. 

Although being designed for the search for nuclear recoils, the CDMS experiment has started to look for electromagnetic signatures of dark matter particles. The analysis of the low-energy electron-recoil spectrum of CDMS may help to identify or constrain possible models which can explain the annual modulation signature observed by DAMA.


\begin{footnotesize}



%

\end{footnotesize}


\end{document}